\documentclass[]{interact}

\usepackage{epstopdf}
\usepackage{subfigure}
\usepackage{float}

\usepackage[numbers,sort&compress,merge]{natbib}
\bibpunct[, ]{(}{)}{,}{n}{,}{,}

\theoremstyle{plain}

\theoremstyle{definition}

\theoremstyle{remark}

\begin{document}
\title{Controlling steady-state second harmonic signal via linear and nonlinear Fano resonances}
\author{
\name{Mehmet G\"{u}nay\textsuperscript{a}\thanks{CONTACT M.~G. Author. Email: gunaymehmt@gmail.com \hspace{0.25cm} Tel:+90 537 669 3386}
, Zafer Artvin\textsuperscript{b,c}\thanks{Z.~A. Author. Email: zartvin@metu.edu.tr \hspace{0.25cm} Tel: +90 553 351 0277},  Alpan Bek\textsuperscript{d}\thanks{A.~B. Author. Email: bek@metu.edu.tr \hspace{0.25 cm} Tel: +90 312 210 5075}  and Mehmet Emre Ta{s}g{i}n\textsuperscript{a}\thanks{CONTACT M.~E.~T. Author. Email: metasgin@hacettepe.edu.tr \hspace{0.25cm} Tel:+90 312 780 6312 }}
\affil{\textsuperscript{a}Institute  of  Nuclear  Sciences, Hacettepe University, 06800 Ankara, Turkey; \textsuperscript{b}Department of Nanotechnology and Nanomedicine, Hacettepe University, 06800 Ankara, Turkey; \textsuperscript{c}Central Laboratory, Middle East Technical University, 06800 Ankara, Turkey; \textsuperscript{d}Department of Physics, Middle East Technical University, 06800 Ankara, Turkey}
}
\maketitle

\begin{abstract}
Nonlinear signal even from a single molecule becomes visible at hot spots of plasmonic nanoparticles. In these structures, Fano resonances can control the nonlinear response in two ways. \textit{(i)} A linear Fano resonance can enhance the hot spot field, resulting enhanced nonlinear signal. \textit{(ii)}   A nonlinear Fano resonance can enhance the nonlinear signal without enhancing the hot spot. In this study, we compare the enhancement of second harmonic signal at the steady-state obtained via these two methods. Since we are interested in the steady-state signal, we adapt a linear enhancement which works at the steady-state. This is different than the dark-hot resonances that appears in the transparency window due to enhanced plasmon lifetime.
\end{abstract}
\section{Introduction}
Metal nanoparticles~(MNPs) trap incident radiation into nm-size hot spots as localized surface plasmon~(LSP) oscillations. Intensity of the hot spot (near field) can be $ 10^5 $ times of the one for the incident field~\cite{hot_spot1,hot_spot2} or even further~\cite{hot_spot3}. This phenomenon enables several technical applications based on the enhancement of linear and nonlinear responses. For instance, a fluorescent molecule in the vicinity of the hot spot becomes detectable~\cite{fluorescence_1,fluorescence_2}, which would not be possible with a weak intensity incident field. Localization also strengthens the nonlinear properties of molecules positioned in the vicinity of the hot spots. Signal from a Raman-reporter molecule can be enhanced remarkably \cite{Raman_reporter}. Such a great increase in the signal results from the localization of both the incident and the produced~(Raman) fields~\cite{SERS}. In this way, surface enhanced Raman scattering~(SERS) enables detection of Raman signal even from a single molecule. Similarly, nonlinear processes like second harmonic generation~(SHG)~\cite{Enh_SHG,Enh_SHG2} and four wave-mixing~(FWM)~\cite{Enh_FWM,Enh_FWM2} are also enhanced.

Enhanced hot spot field also enables the observation of Fano resonances~\cite{Fano_observation1,Fano_observation2}, analog of electromagnetically induced transparency~(EIT)~\cite{SC}. In EIT, excited level is hybridized by coupling to a third one via a strong microwave field. Hybridization is weak such that the splitting in the excited state remains in the spectral width of the excited state. The two paths operate out of phase, one emits while the other one absorbs, and a transparency window appears. A transparency window, Fano resonance~(FR), also appears in the plasmon response, when a quantum emitter~(QE) is placed in the hot spot of MNP~\cite{hot_spot1}. In a FR, the weak hybridization is induced by the interaction of the QE with the own polarization of the plasmon near field. A FR can appear when the plasmon mode of MNP is coupled to a longer lifetime excitation. For instance, it was shown that FR can appear in all-plasmonic system when the excited LSP is coupled to a long living dark plasmon mode~\cite{Tassin,Liu}. 

FRs can control linear and nonlinear properties of MNPs~\cite{FR_control1,FR_control2,FR_control3,FR_control4,FR_control5}. Typical lifetime of LSP oscillations is extremely short, i.e. $ 10^{-14}-10^{-13}  $ seconds. A FR can increase the lifetime of plasmon excitation~\cite{life_1,life_2,lifex,Bilge_thesis}. In the same incident field strength, a MNP with a FR can accumulate more intense field at the hot spot. These are called as dark-hot resonances~\cite{dark_hot}. This enhancement in the plasmon lifetime enables the operation of spasers~\cite{spaser,spaser2}, nano-lasers. Dark-hot resonances appear strongest when the incident field is resonant to the transparency window~\cite{dark_hot}. Even though excitation reaches the transparency (zero) in the steady state, hot spot becomes more intense for some time due to enhanced plasmon lifetime. Dark-hot resonances are cleverly adapted in enhancing the SERS~\cite{dark_hot_SERS1,dark_hot_SERS2,dark_hot_SERS3} and FWM~\cite{dark_hot_FWM} signals further. For instance, a double resonance system, where exciting and converted frequencies are tuned to two plasmon resonances~\cite{double_resonant}, can yield a stronger SERS signal. Besides that, when exciting and converted frequencies are aligned with two FRs, then the signal can be enhanced several orders on the top of the double resonance scheme. One can conduct similar results also for SHG process. 

In this manuscript, we study the control of steady-state of second harmonic signal via path interferences (\textit{i}) in the linear and (\textit{ii}) nonlinear responses. For enhancing the linear response (\textit{i}), so the nonlinear one, we use path interference scheme different than the dark-hot resonances. The scheme we use, a novel scheme to our best knowledge, enhances the hot spot field intensity in the steady-state, which also leads to enhancement in the second harmonic field intensity. Next, we use a path interference scheme~\cite{Raman_reporter} which enhances only (\textit{ii}) the nonlinear response without increasing the hot spot field intensity. This method can be useful when MNP-QE hybrid system is excited with a strong pulsed laser. For instance, experiments using a molecule type with very weak SHG response would necessitate a strong laser. This may heat the molecule too much in the presence of a dark-hot resonance or enhancement via method (\textit{i}) we introduced here. In such a case, enhancing the second harmonic signal without increasing the hot spot field intensity could be very useful. 

It is demonstrated that all plasmonic FRs can silence SHG. Here, we show that such a silencing~(suppression) is also possible in the presence of a molecule resonant to the second harmonic signal. We use a Lorentzian dielectric function for molecule and demonstrate the suppression with the exact solution of 3D Maxwell equations. This was predicted by an analytical method~\cite{Enh_SHG2} previously but not confirmed with 3D solutions.

\begin{figure}
{\resizebox*{14.3cm}{!}{\includegraphics{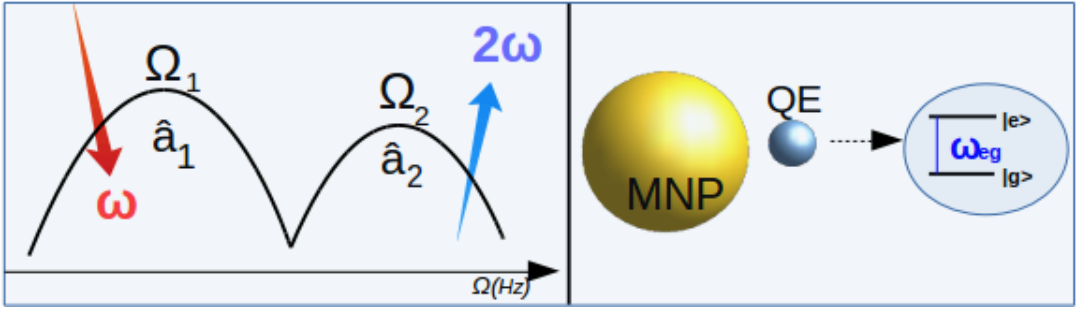}} }
\caption{ \textit{Left.} Schematic description of the two plasmon modes, $\hat{a}_1$ and  $\hat{a}_2$ involved in the SHG process~\cite{Grosse}. Incident field of frequency $\omega$ excites $\hat{a}_1$ mode. At the hot spot, two $\omega$ plasmons combine to yield a single $2\omega$ plasmon in the $\hat{a}_2$ mode. SHG process takes place in between plasmons~\cite{Grosse} since the overlap integral in Eq.(\ref{HamSh}) becomes largest for both $\omega$ and $2\omega$ are localized. \textit{Right.} A QE of level spacing $\omega_{eg}$ can be chosen to couple either \textit{(i)} $\hat{a}_1$ mode, $\omega_{eg}\sim \Omega_1$ or \textit{(ii)} to $\hat{a}_2$ mode, $\omega_{eg}\sim \Omega_2$. Different path interference schemes result in the two cases, see Figure~\ref{Coupled_to_Linear} and Figure~\ref{Coupled_to_SH}.} \label{Sketch}
\end{figure}

The manuscript is organised as follows. In Sec.2, we describe the mechanism for the SHG processes in plasmonic nano-materials. We derive effective Hamiltonian for a second harmonic converter MNP coupled with a QE. We derive the equations of motion for the plasmon modes and the QE. In Sec.3, we discuss two different enhancement schemes. In Sec.3.1, we choose the level spacing~($\omega_{eg}$) of the QE such that it is coupled to the driven plasmon mode, $\hat{a}_1$ in Figure~\ref{Sketch}. This enhances the steady-state intensity of the hot spot, so the second harmonic signal which is proportional to the square of the hot spot field. In Sec.3.2, we couple the QE to the $\hat{a}_2$ mode in which second harmonic oscillations take place. This enhances the second harmonic signal without altering the hot spot field intensity of the driven $\hat{a}_1$ mode. Sec.4 contains our conclusions.   

\section{Dynamics of the system} 
In this section, we derive the effective Hamiltonian for a plasmonic second harmonic converter coupled to a QE. Hamiltonian contains two possibilities; QE coupled to \textit{(i)} driven $\hat{a}_1$ mode and to the \textit{(ii)} high energy plasmon mode $\hat{a}_2$ into which second harmonic oscillations take place. We derive the equations of motion for each case. These equations are the basis on which we discuss how one can gain control over the nonlinear signal with and without modifying linear response.  

\subsection{Hamiltonian} 

SHG~(frequency doubling) process in a plasmonic nano-material can be described as follows. When plasmonic structure is illuminated with a strong incident laser field ($ \varepsilon_p $) of frequency $ \omega $, the plasmon oscillations are excited in the $\hat{a}_1$ mode. Combination of the two plasmon in the $ \hat{a}_1 $ mode yield to create a single plasmon in the $ \hat{a}_2 $ mode~\cite{Finazzi} oscillating as $ e^{-i2\omega t} $. Second harmonic conversion process takes place between plasmons~\cite{Grosse,FR_control2,Wunderlich}  since the overlap integral, Eq.\ref{HamSh} being, attains a large value.
The Hamiltonian of this process can be written as 
\begin{eqnarray}
 \hat{H}_{sh}=\Bigl\{\int d^3r \chi_{2}(\textbf{r}) E_2^\ast(\textbf{r}) E_1^2(\textbf{r})\Bigr\} \hat{a}_2^\dagger \hat{a}_1 \hat{a}_1 +\it{hc} .
\label{HamSh}
\end{eqnarray}
The integral in the parenthesis, $ \chi^{(2)}= \int d^3r \chi_{2}(\textbf{r}) E_2^\ast(\textbf{r}) E_1^2(\textbf{r})$, is the overlap integral. It (\textit{i}) determines the strength of the SHG process, (\textit{ii})  demonstrates us why second harmonic conversion takes place between plasmon of different frequencies and (\textit{iii}) gives the selection rules for SHG process. 
Why SHG is not observed in the far-field for a centro-symmetric plasmonic nano-structure can be observed from  this overlap integral $ \chi^{(2)} $ ~\cite{Tasgin}. When  a QE couples to one of the hot spots of the MNP, it strongly interacts with that mode. Small decay rate of QE creates a weak hybridization, in which two or more absorption/emission paths come into play. Depending on the level spacing ($\omega_{eg}$) of the QE, it is possible to introduce different interference schemes.

The Hamiltonian of the system can be written as the sum of the  energy of the plasmon oscillations~($\Omega_1,\Omega_2  $), quantum emitter~($  \omega_{eg}$) and the energy transferred by the pump source~($\omega $)
\begin{eqnarray}
\hat{H}_0&=&\hbar \Omega_1 \hat{a}_1^\dagger \hat{a}_1+\hbar \Omega_2 \hat{a}_{2}^\dagger \hat{a}_2+\hbar \omega_{eg} |e \rangle \langle e|\\
\hat{H}_{P}&=&i\hbar (\varepsilon_p  \hat{a}_1^\dagger e^{-iwt} -\textit{h.c})
\label{Ham0}
\end{eqnarray} 
as well as the SHG process $\hat{H}_{sh} $ as defined in Eq.(\ref{HamSh}) and the interaction of the QE with the plasmon-polariton modes  $ \hat{H}_{int}$.
\begin{eqnarray}
\hat{H}_{sh} &=& \hbar \chi^{(2)} (\hat{a}_2^\dagger \hat{a}_1 \hat{a}_1+ \hat{a}_1^\dagger \hat{a}_1^\dagger \hat{a}_2 )\\  
\hat{H}_{int}&=&\hbar (f_1 \hat{a}_1^\dagger+f_2\hat{a}_2^\dagger) |g\rangle \langle e|+ \textit{h.c} 
\end{eqnarray}
Here the parameters  $ f_1 $ and $ f_2 $ , in units of frequency, are the coupling strengths of the linear and second harmonic responses of the MNP to the QE respectively. $ \chi^{(2)} $ is the overlap integral defined in Eq.(\ref{HamSh}) and  $|g\rangle$~($|e \rangle$) is the ground~(excited) state of the QE.
\subsection{Equations of motion} 
The equations of the motion  of the system can be derived by using Heisenberg equations (e.g. $i\hbar\hat{a}_i=[\hat{a}_i,\hat{H}] $). In this work, we are not interested in the quantum optical features of the modes. We replace the operators $\hat{a}_i$ and $ \hat{\rho}_{ij}= |i \rangle \langle j|$ with  complex numbers  ${\alpha}_i$ and $ {\rho}_{ij} $~\cite{Premaratne2017} respectively and equations of the motion for the plasmon amplitudes and density matrix elements, respectively, can be obtained as
\begin{subequations}
\begin{align}
\dot{{\alpha}}_1&=-(i\Omega_1+\gamma_{1}) {\alpha}_1-i 2\chi^{(2)} {\alpha}_1^\ast {\alpha}_2-i f_1{{\rho}}_{ge}+\varepsilon_p e^{-iw t} \label{EOMa},\\
\dot{{\alpha}}_2&=-(i\Omega_2+\gamma_{2}) {\alpha}_2-i \chi^{(2)} {\alpha}_1^2-i f_2 {\tilde{\rho}}_{ge} \label{EOMb},\\
\dot{{\rho}}_{ge} &=  -(i \omega_{eg}+\gamma_{eg}) {\rho}_{ge}+i (f_1 {\alpha}_1+f_2{\alpha}_2)({\rho}_{ee}-{{\rho}}_{gg}) \label{EOMc},\\
\dot{{{\rho}}}_{ee} &= -\gamma_{ee} {{\rho}}_{ee}+i \bigl \{(f_1 {\alpha}^\ast_1+f_2 {\alpha}^\ast_2) {{\rho}}_{ge}- \textit{c.c} \bigr \}
\label{EOMd},
\end{align}
\end{subequations}
where  $ \gamma_{1,2} $ and $ \gamma_{ee} $ are the damping rates  of plasmon modes and quantum emitter, respectively. The  conservation of probability  ${\rho}_{ee}+{\rho}_{gg} =1$ and population inversion $y={\rho}_{ee}-{\rho}_{gg}$ with the off-diagonal decay rate of the QE  $ \gamma_{ge} = \gamma_{ee}/2 $ accompanies Eqs.(\ref{EOMa}-\ref{EOMd}). 

In the steady state, the two plasmon mode amplitudes and the diagonal density matrix elements can be written as
\begin{eqnarray}
{\alpha}_1(t)= \tilde{\alpha}_1e^{-i \omega t}, \qquad {\alpha}_2(t)=\tilde{\alpha}_2 e^{-i2 \omega t}, \qquad {\rho}_{ee}(t)=\tilde{\rho}_{ee}, \qquad {\rho}_{gg}(t)=\tilde{\rho}_{gg}.
\label{steady}
\end{eqnarray}
The off-diagonal density matrix element $\rho_{ge}(t)$ determines the electric polarization of the QE. Its steady-state oscillations take the form  $\rho_{ge}(t)=\tilde{\rho}_{ge} e^{-i\omega t}$ and $\rho_{ge}(t)=\tilde{\rho}_{ge} e^{-i2\omega t}$ when QE is coupled to $\alpha_1$ mode and $\alpha_2$ mode respectively. Although, we present our results with the numerical time evolutions of Eqs.(\ref{EOMa}-\ref{EOMd}) in Figure~\ref{Coupled_to_Linear} and Figure~\ref{Coupled_to_SH}, we study the steady-state carefully in order to gain understanding over different interference schemes. 

\section{Enhancement and suppression}
In this section, we  examine the  second harmonic response of the plasmonic nanostructure  coupled to a QE. The relative enhancement factor~(EF) for each mode is defined in the presence and absence of a QE as
\begin{eqnarray}
 EF_{\tilde{\alpha}_1}= \frac{|\tilde{\alpha}_1(f_i\neq0)|^2}{|\tilde{\alpha}_1 (f_i=0)|^2}, \qquad
 EF_{\tilde{\alpha}_2}= \frac{|\tilde{\alpha}_2(f_i\neq0|^2}{|\tilde{\alpha}_2 (f_i=0)|^2}, \qquad i=1,2.
\label{EF}
\end{eqnarray}
 We aim to control the amplitudes of both linear and nonlinear fields. In order to do that, we use a basic model and obtain the steady-state amplitudes for the generated second harmonic plasmons. From a single equation we demonstrate the origin of  enhancement and suppression. We compare two different cases for the level spacing of the QE  close to the  (\textit{i}) linear and  (\textit{ii}) second harmonic frequency  oscillations. We show that in either case enhancement and suppression of the SHG process is possible. In the first case both enhancement and suppression in the second harmonic field  is proportional to square of the linear field. In the latter case, however, the enhancement can be obtained  \textit{silently} i.e without modifying hot spot field intensity of  $ \alpha_1 $ mode. 
Then, we compare the predictions of our model in suppression phenomena with 3-dimensional FDTD simulations. We confirm that suppression  of the SHG process is observed in the predicted spectral position.

\subsection{Coupling quantum emitter to $ \alpha_1 $ mode}

Illumination of MNPs with an electromagnetic field of frequency $\omega$ results in strong field enhancement in the vicinity of the MNPs. By placing QE into these regions, further enhancement in the hot spot field intensities, due to linear Fano resonances, can be obtained. In this part, we show that hot spot field intensity can be enhanced not only in the initial times, but also in the steady-state when a QE is coupled to driven mode of the MNP. We also demonstrate the cosequences of this process in the nonlinear response. 

\begin{figure}[H]
\centering
{
\resizebox*{19cm}{!}{\includegraphics{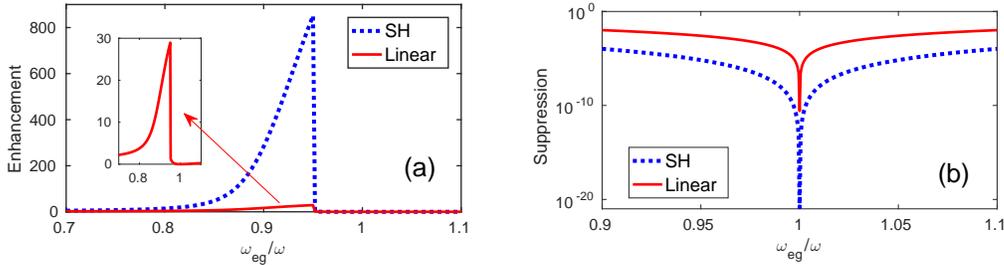}}}\hspace{5pt}
\caption{(a) Enhancement  and (b) suppression of the second harmonic (blue-dotted) and linear (red-straight) field intensities for $ \Omega_1=0.8\omega $ when QE coupled to $\tilde{\alpha}_1$ mode, obtained from time evolution of Eqs.(\ref{EOMa})-(\ref{EOMd}). We use $ \chi^{(2)} =10^{-5}\omega$, $f_1 = 0.1\omega$, $f_2 = 0$, $ \gamma_{1}=\gamma_{2}=0.01\omega $ and $ \gamma_{eg}=10^{-5}\omega $.  }
\label{Coupled_to_Linear}
\end{figure}

The enhancement and suppression of the second harmonic field intensity is possible by enhancing the linear response. When a QE is couple to $ \alpha_1 $ mode, path interference effects  take place for the linear field intensity. Since we are examining the coupling of QE to the driven mode, interaction strength between QE and second harmonic response can be neglected (e.g. $ f_2\approx0 $). The source term for off-diagonal density matrix element, however, can be considered as $ \alpha_1 $ mode. Inserting $ {\rho}_{ge}= \tilde{\rho}_{ge} e^{-i \omega t} $ and Eq.(\ref{steady}) into Eqs.(\ref{EOMa}-\ref{EOMd}), one can find steady state solution for $\tilde{\alpha}_1$ and $\tilde{\alpha}_2$ modes as
\begin{eqnarray}
\label{a1_coupling}
\tilde{\alpha}_1&=&\frac{\varepsilon_p }{[i(\Omega_1- \omega)+\gamma_1]-\frac{f_1^2y}{[i(\omega_{eg}-\omega)+\gamma_{eg}]}}+f(\chi^{(2)}) \\ 
 \tilde{\alpha}_2&=&\frac{-i \chi^{(2)}}{[i(\Omega_2- 2\omega)+\gamma_2]} \tilde{\alpha}_1^2 . 
   \label{a2_mode}
\end{eqnarray}
The enhancement of $\tilde{\alpha}_1$ intensity can be obtained by minimizing the denominator of the Eq.(\ref{a1_coupling}). When oscillation frequency of the linear response, $ \Omega_1 $, is off-resonant with $\omega $, the residual part of the first term can be eliminated by choosing appropriately level spaced~($ \omega_{eg} $) QE and/or tuning interaction strength~($ f_1 $) by adjusting separation between MNP and QE. That is, for the choice of the level spacing, $ \omega_{eg}\simeq \omega-\frac{f_1^2 y}{\Omega_1-\omega} $, which cancels imaginary part of the first term, the hot spot field intensity can be enhanced about 30 times. Due to quadratic dependece on $\tilde{\alpha}_1$ in Eq.(\ref{a2_mode}), the enhancement in the second harmonic is approximately square of the linear mode, see Figure~\ref{Coupled_to_Linear}a.

When $\omega= \omega_{eg} $ and $f_1$ is significant, the second term in the denominator of the Eq.(\ref{a1_coupling}) becomes ${f_1^2y}/{\gamma_{eg}}$. When we scale all the frequencies by the optical drive, $\omega$, $\gamma_{eg}^{-1}$ becomes very large. In this case, the second term dominates the denominator and results a very small $\tilde{\alpha}_1$ which yields the observed transparency for linear field and hence for the nonlinear field. This is depicted in Figure~\ref{Coupled_to_Linear}b. 

\subsection{Coupling quantum emitter to $ \alpha_2 $ mode}
SHG processes can be enhanced and suppressed by constructive and destructive interferences of the frequency conversion paths. If the level spacing of QE is chosen closely to second harmonic frequency $ \omega_{eg}\approx 2\omega $ its interaction with the $ \alpha_2 $ mode becomes stronger compared to $ \alpha_1 $ mode as it is highly off-resonant from driven mode. In addition, the hot spots of each mode can emerge at different spatial positions. By placing QE at the hot spot of $ \alpha_2 $ mode,  coupling of  QE to the driven mode becomes negligible (i.e. $ f_1\approx 0 $). One can observe that steady-state solution for the QE is in the form  $ {\rho}_{ge}(t)= \tilde{\rho}_{ge} e^{-i2 \omega t}$ with $\tilde{\rho}_{ge} $ is a constant in time. Inserting this form into Eq.(\ref{EOMc}) and using Eq.(\ref{steady}), the steady state amplitude for the plasmons in the $\alpha_2$ mode can be obtained  from Eqs.(\ref{EOMa}-\ref{EOMd}) as
\begin{eqnarray}
\tilde{\alpha}_2=\frac{-i \chi^{(2)}\tilde{\alpha}_1^2 }{[i(\Omega_2- 2\omega)+\gamma_2]-\frac{f_2^2y}{[i(\omega_{eg}-2 \omega)+\gamma_{eg}]}}.  \label{SH_amp}
\end{eqnarray}
Interpretation  of why enhancement and suppression take place in second harmonic field can be done  by  examining the denominator of the Eq.(\ref{SH_amp}). Here, we demonstrate how second harmonic field can be enhanced without any modification in the linear response, $\tilde{\alpha}_1$.

\begin{figure}
\centering
{
\resizebox*{18cm}{!}{\includegraphics{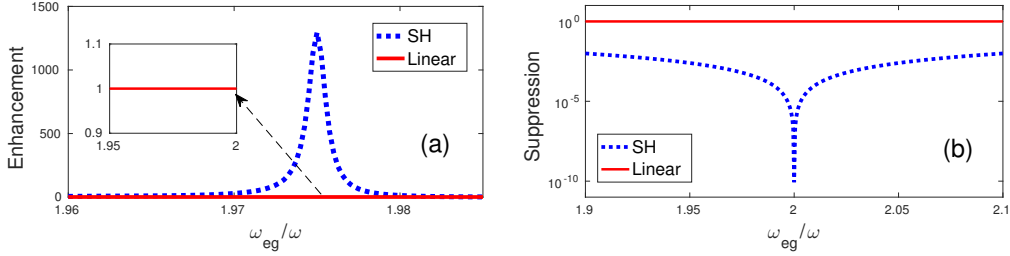}}}\hspace{5pt}
\caption{(a) Enhancement and (b) suppression of the second harmonic (blue-dotted) and linear (red-straight) field intensities when QE coupled to $ \alpha_2 $ mode, obtained from time evolution of Eqs.(\ref{EOMa})-(\ref{EOMd}). The QE is chosen to have no SH response. Both enhancement and suppression factors are calculated with using Eq.(~\ref{EF}). The same parameters are used as in Figure~\ref{Coupled_to_SH}.} \label{Coupled_to_SH}
\end{figure}

Similar to previous part, the enhancement of $\tilde{\alpha}_2$ intensity can be obtained by minimizing the denominator of the Eq.(\ref{SH_amp}). That is, for the choice of the level spacing, $ \omega_{eg}\simeq 2\omega-\frac{f_2^2 y}{\Omega_2-2\omega} $, which cancels imaginary part of the first term, second harmonic field intensity can be enhanced about 1500 times without changing linear hot spot field intensity, see Figure~\ref{Coupled_to_SH}a. The enhancement factor in Figure~\ref{Coupled_to_SH}a is obtained for the choice of $ \Omega_2=1.6\omega $ and the interaction strength  $f_2=0.1\omega  $. For a high quality plasmonic nanostructure~\cite{smallgamma}, this factor can be even larger.

The denominator of the Eq.(\ref{SH_amp}) also demonstrates us the possibility  of a SHG suppression effect. We examine the denominator of the last term in the denominator of the Eq.(\ref{SH_amp}). When a QE level spacing is chosen close to the second harmonic frequency $ \omega_{eg} \approx 2 \omega $, the last term becomes $ {f_2^2y}/{\gamma_{eg}} $. Typical values for the decay rates of QEs are very small compared to all other frequencies. For example $ \gamma_{eg}\sim 10^{-6}\omega $ for quantum dots~\cite{Kosionis}. With a coupling constant of sufficient strength, $ {f_2^2y}/{\gamma_{eg}} $ term dominates the denominator of the Eq.(\ref{SH_amp}) and suppression of the second harmonic field emerges, see Figure~\ref{Coupled_to_SH}b. Here, again we do not observe any change in the linear field intensity.

We underline that the results presented in Figure~\ref{Coupled_to_Linear} and Figure~\ref{Coupled_to_SH} are the exact solutions of Eqs.(\ref{EOMa}-\ref{EOMd}) without any approximation.

We note an important point. For the system parameters we treat as a sample system, (\textit{i}) enhancement of the linear response, i.e. $\tilde{\alpha}_1$, appears rather small, about 30 times. When we compare this with the (\textit{ii}) nonlinear Fano resonance method, SHG enhancement from (\textit{ii})  appears larger. However, this observation can change dramatically for using a sample system with different parameters. Enhancement via linear Fano resonances, in general, can be expected to be much larger than the nonlinear Fano resonances. Because SHG enhancement with linear Fano resonance is the square of the linear response, which can be more than two orders of magnitude in other systems.

\subsection{Comparision of the suppression in 3D simulations}
Suppression of the SHG with all-plasmonic Fano resonances was shown in Ref.~\cite{FR_control3}. In a previous study~\cite{Enh_SHG2}, we anticipated the presence of the suppression effect also for a Fano resonance originating from a QE. The method we have used in this work~\cite{Enh_SHG2} was a first order approximation for SHG. Here, we use self consistent Maxwell equations also for the SHG process and demonstrate the presence of the anticipated suppression effect with 3D-FDTD simulations.

\begin{figure}
\centering
\subfigure[Electric field intensity in the presence (blue-straight) and absence (black-dotted) of QE having a sharp resonance at $ \lambda_{QE}= $ 530 nm. The insets magnify the linear~(top) and the second harmonic~(bottom) field intensity regions.]{
\resizebox*{6.9cm}{!}{\includegraphics{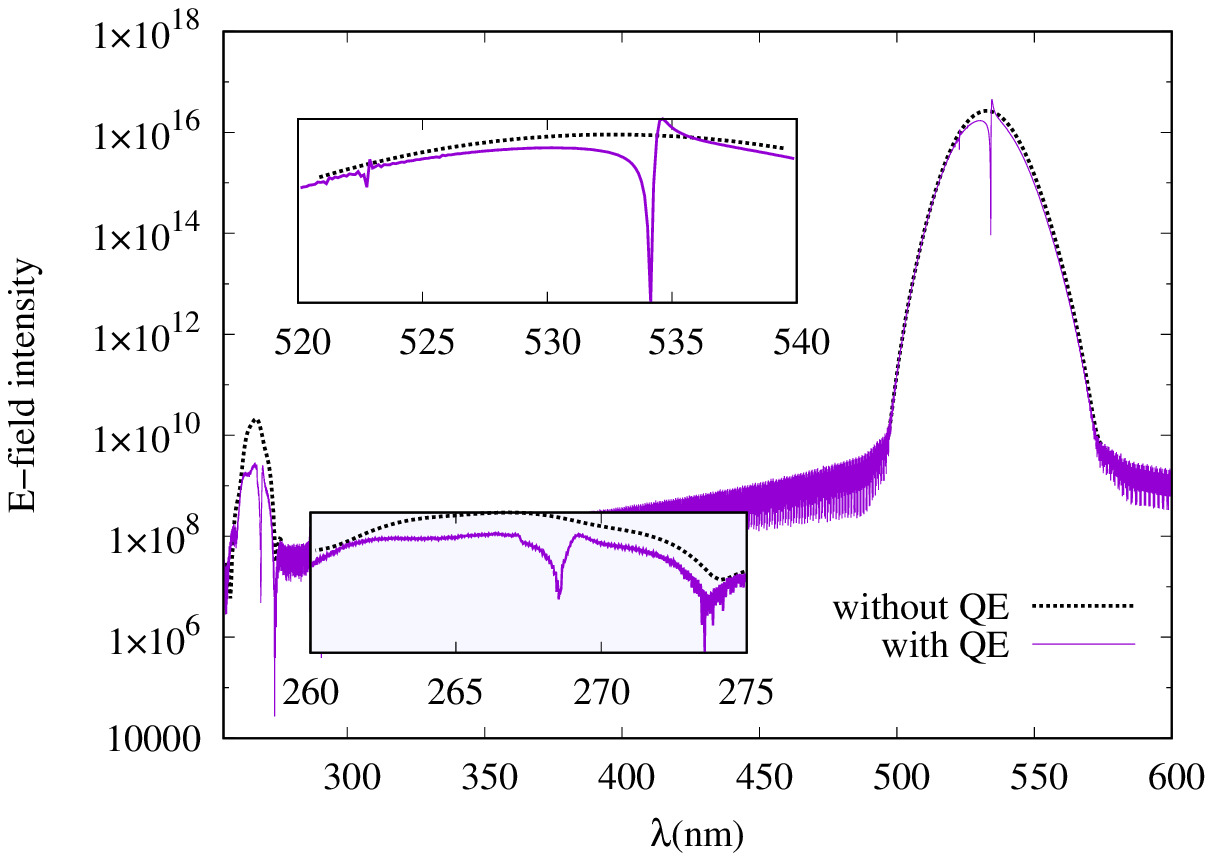}}}\hspace{10pt}
 \subfigure[Electric field intensity in the presence (blue-straight) and absence (black-dotted) of QE having a sharp resonance at $ \lambda_{QE}= $ 265 nm. The inset magnifies the second harmonic field intensity region.]{
\resizebox*{6.7cm}{!}{\includegraphics{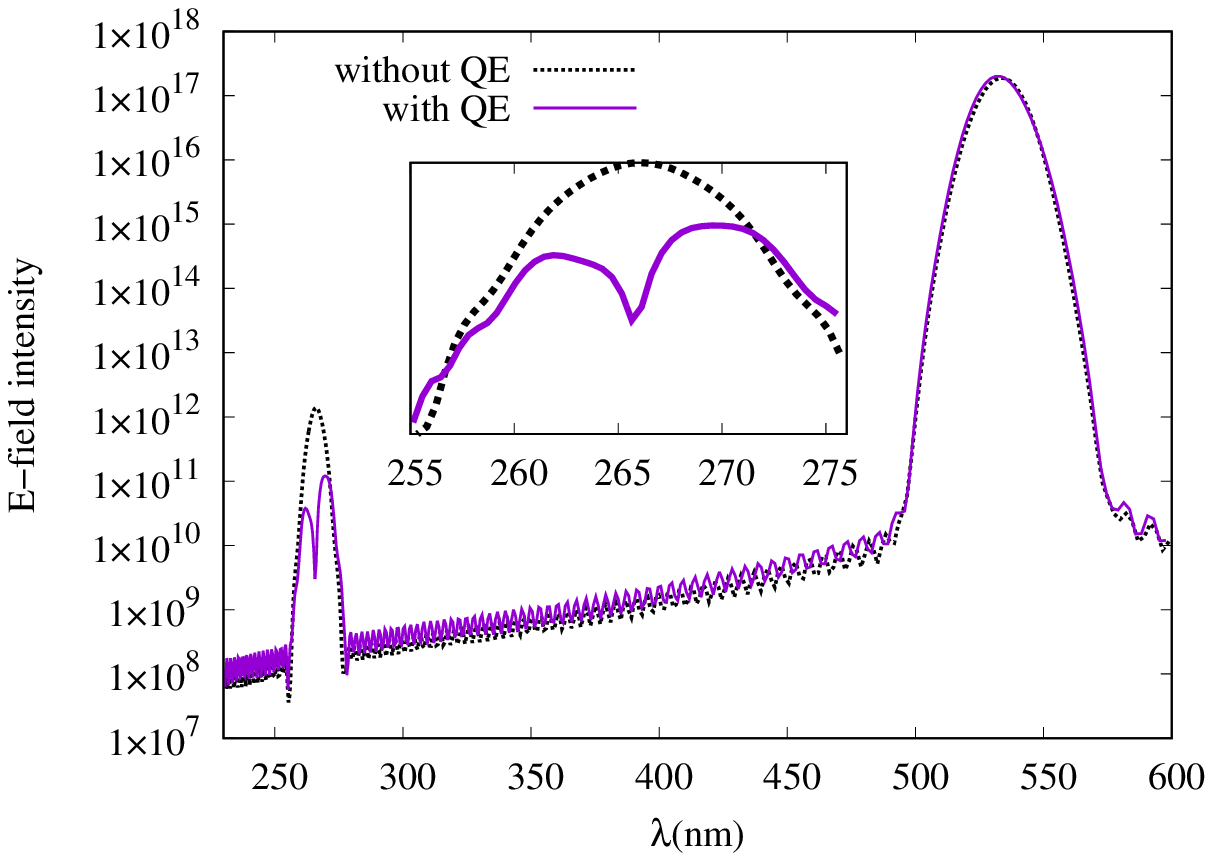}}}
\caption{Suppression effect when QE is coupled to (a) $ \alpha_1 $ and (b) $ \alpha_2 $ modes of MNP, obtained from FDTD simulation.} \label{FDTD}
\end{figure}

In Figure~\ref{FDTD}, we compare the predictions of our model for suppression effect with the 3D FDTD simulation in which  Maxwells equations can be solved using conventional Yee cell algorithm~\cite{Yee}. We use nanospheres for MNP and QE as in Figure~\ref{Sketch}. Radii of the gold nanoparticle and QE are 35 nm and 15 nm respectively. We use experimental data for the dielectric function of the gold nanoparticle~\cite{gold_dielectric} with $\chi^{(2)}$ nonlinearity. We place an auxilary QE in the vicinity of the MNP, which has a Lorentzian dielectric function. The Lorentzian peak is centred around 530 nm for Figure~\ref{FDTD}a and 265 nm for Figure~\ref{FDTD}b. Excitation wavelength is $  \lambda_{exc}=$ 530 nm with 30 THz bandwidth and  amplitude of the source beam is set $ 10^6$ V/m to have a strong nonlinear response. In Figure~\ref{FDTD}a driven mode of MNP strongly interacts with QE having a sharp resonance at $ \lambda_{QE}= $ 530 nm, suppression emerges in both modes, see insets of Figure~\ref{FDTD}a. When a gold-MNP strongly interacts with QE having a sharp resonance at $ \lambda_{QE}=$ 265 nm, the suppression in the second harmonic response $ \lambda_{SH} $ emerges without significant change in the linear response, see Figure~\ref{FDTD}b.   We wanted to be confident about the origin of the dip at 265 nm in Figure~\ref{FDTD}. In order to remove the possibility that dip originates from the absorption induced by QE, we illuminated the sample by a pulse centered at 265 nm and checked if a linear resonance appears. Since we observed that Fano resonance, we became confident that dip does not originate from absorption.

\section{Conclusions} 
We demonstrate that one can gain control over the steady-state second harmonic signal from a plasmonic nano-material via two kinds of Fano resonances. In the first method, with linear Fano resonance, one can increase the hot spot field intensity for the driven field, $ \sim e^{-i \omega t} $. This also increases the second harmonic signal, $ P_{2 \omega}\sim E_\omega^2 $. We develop a method for enhancing the steady state linear response of a MNP. This is different than the dark-hot resonances which appear at the transparency window due to plasmon lifetime enhancement. In the second method, with nonlinear Fano resonance, one can gain control over the second harmonic signal without altering the hot spot field intensity for the driven mode. Using the exact solutions of Maxwell equations, we also show that second harmonic signal is suppressed when the QE is tuned to $\omega_{eg}=2\omega$, as predicted~\cite{Enh_SHG2}. This phenomenon has also been observed in all-plasmonic Fano resonances~\cite{FR_control3,Bilge}.

\section*{Acknowledgments}
MG and MET acknowledges support from TUBITAK Grant No. 117F118.

\end{document}